\documentclass[%
 aip,
 amsmath,amssymb,
 reprint,%
]{revtex4-2}

\usepackage{graphicx}
\usepackage{dcolumn}
\usepackage{bm}

\usepackage[utf8]{inputenc}
\usepackage[T1]{fontenc}
\usepackage{mathptmx}
\usepackage{etoolbox}
\usepackage{hyperref}
\usepackage{float}

\makeatletter
\def\@email#1#2{%
 \endgroup
 \patchcmd{\titleblock@produce}
  {\frontmatter@RRAPformat}
  {\frontmatter@RRAPformat{\produce@RRAP{*#1\href{mailto:#2}{#2}}}\frontmatter@RRAPformat}
  {}{}
}%
\makeatother
\begin{document}

\preprint{AIP/123-QED}

\title[Imaging dynamic exciton interactions and coupling in transition metal dichalcogenides]{Imaging dynamic exciton interactions and coupling in transition metal dichalcogenides}
\author{Torben L. Purz}
\affiliation{ 
Department of Physics, University of Michigan, Ann Arbor MI 48109-1040, USA
}%
\author{Eric W. Martin}%
\affiliation{ 
MONSTR Sense Technologies LLC, Ann Abor MI 48104, USA
}%
 \author{William G. Holtzmann}
 \affiliation{ 
Department of Physics, University of Washington, Seattle, WA 98195-1560, USA
}%
 \author{Pasqual Rivera}
 \affiliation{ 
Department of Physics, University of Washington, Seattle, WA 98195-1560, USA
}%
 \author{Adam Alfrey}
 \affiliation{ 
Department of Physics, University of Michigan, Ann Arbor MI 48109-1040, USA
}%
\author{Kelsey M. Bates}
\affiliation{ 
Department of Physics, University of Michigan, Ann Arbor MI 48109-1040, USA
}%
 \author{Hui Deng}
 \affiliation{ 
Department of Physics, University of Michigan, Ann Arbor MI 48109-1040, USA
}%
 \author{Xiaodong Xu}
 \affiliation{ 
Department of Physics, University of Washington, Seattle, WA 98195-1560, USA
}%
 \author{Steven T. Cundiff}
 \email{cundiff@umich.edu}
\affiliation{ 
Department of Physics, University of Michigan, Ann Arbor MI 48109-1040, USA
}%

\date{\today}

\begin{abstract}
Transition metal dichalcogenides (TMDs) are regarded as a possible materials platform for quantum information science and related device applications. In TMD monolayers, the dephasing time and inhomogeneity are crucial parameters for any quantum information application. In TMD heterostructures, coupling strength and interlayer exciton lifetimes are also parameters of interest.
However, many demonstrations in TMDs can only be realized at specific spots on the sample, presenting a challenge to the scalability of these applications. Here, using multi-dimensional coherent imaging spectroscopy (MDCIS), we shed light on the underlying physics - including dephasing, inhomogeneity, and strain - for a MoSe\textsubscript{2} monolayer and identify both promising and unfavorable areas for quantum information applications. We furthermore apply the same technique to a MoSe\textsubscript{2}/WSe\textsubscript{2} heterostructure. Despite the notable presence of strain and dielectric environment changes, coherent and incoherent coupling, as well as interlayer exciton lifetimes are mostly robust across the sample. This uniformity is despite a significantly inhomogeneous interlayer exciton photoluminescence distribution that suggests a bad sample for device applications.
This robustness strengthens the case for TMDs as a next-generation materials platform in quantum information science and beyond.
\end{abstract}

\maketitle

\section{Introduction}

Transition metal dichalcogenides (TMDs) are regarded as a prime materials platform for applications ranging from solar-energy \cite{TMD_Photodiode,Photovoltaics} and lasers \cite{TMD_Laser} to quantum light-emitting diodes \cite{TMDQuantumLED}.
Especially the rapid charge transfer and associated interlayer excitons \cite{CoherentCouplingPurz,ChargeTransfer,MDCS_CT,ValleySpin, PasqualILE,GalanILE,ILE_review,ILE_Quantum} have received considerable attention. Interlayer excitons with nanosecond lifetimes \cite{ILE_lifetime1,ILE_lifetime2,ILE_lifetime3} that are highly tunable by twist-angle \cite{ILE_twistAngle_lifetime} are potential candidates for qubits \cite{ILE_qubit1,ILE_qubit2}. 
In recent years, strain-engineering in TMD monolayers has also gained momentum with the potential for room-temperature entangled-photon sources\cite{Galan_Wse2,WSe2Pillar_Parag}.
Moreover, coherent coupling between excitons and trions in TMD monolayers \cite{CoherentCoupling,CoherentGalan} and intralayer excitons in TMD heterostructures \cite{CoherentCouplingPurz} has been demonstrated recently, which opens the avenue for quantum coherent control of these materials.

Depending on the sample system, different physical parameters determine the feasibility of quantum information applications. In TMD monolayers, low inhomogeneity and long intralayer exciton dephasing times are crucial. Martin \textit{et al.} demonstrated picosecond dephasing times for MoSe\textsubscript{2} monolayers \cite{Eric_TMD} by measuring the homogeneous linewidth, which is inversely proportional to the dephasing time. These dephasing times are well above previously reported values \cite{PL_LW1,PL_LW2,PL_LW3,PL_LW4} due to the dominant contribution of inhomogeneous broadening in these samples. Meanwhile, Jakubczyk \textit{et al.} \cite{Kasprzak_LW1,Kasprzak_LW2} and Boule \textit{et al.} \cite{Kasprzak_Coherent} have shown broader homogeneous linewidths, with the exciton optical response being in the homogeneous limit for certain areas. As established previously\cite{Kasprzak_LW1,Kasprzak_LW2}, there is an expected anti-correlation between homogeneous and inhomogeneous linewidths, favoring long dephasing times in areas of large inhomogeneity. Non-radiative broadening mechanisms can weaken this anti-correlation.
However, a large inhomogeneity is strongly detrimental to quantum information applications, since it corresponds to several different emitters within the excitation volume. To identify promising areas for these applications, one thus needs to find uniquely suited sample areas with small inhomogeneity and large dephasing times.

\begin{figure*}[t]
\centering
\includegraphics[width=1.0\textwidth]{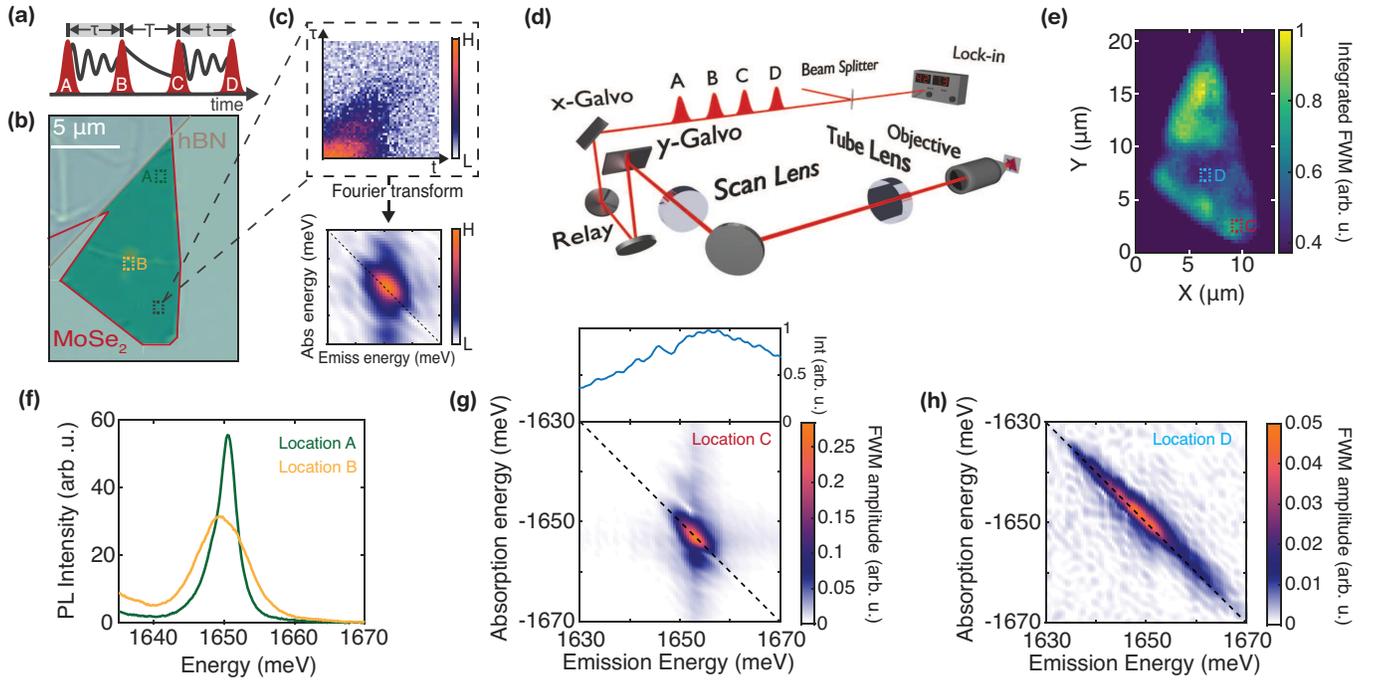}
\caption{\label{fig:Fig1} \textbf{(a)} Schematic of a three-pulse MDCS experiment (fourth
pulse used for heterodyne detection). \textbf{(b)} White light microscopy image of the MoSe\textsubscript{2} monolayer. \textbf{(c)} Schematic of the MDCIS scheme. An image is taken for a fixed $\tau$, $T$, and $t$ position and the stage delays are subsequently varied. Fourier transform yields an MDCS spectrum for every image pixel. \textbf{(d)} Schematic of the custom-built laser scanning microscopy setup. \textbf{(e)} Integrated FWM of the hBN-encapsulated MoSe\textsubscript{2} monolayer. \textbf{(f)} PL-spectrum on the center (yellow) and edge (green) area of the MoSe\textsubscript{2} monolayer. \textbf{(g)} Low-power, low-temperature MDCS spectrum in the bright region at the bottom of the sample (red square). The excitation laser spectrum used for the experiments on the monolayer is shown atop. \textbf{(h)}  Low-power, low-temperature MDCS spectrum in the dark region of the MoSe\textsubscript{2} monolayer towards the center of the sample (blue square).}
\end{figure*}

For heterostructure systems, these considerations change based on the physical effects harnessed for quantum information applications. Interlayer excitons have been proposed as potential candidates for qubits in the past with their formation closely tied to the rapid charge transfer. Moreover, coherent coupling between excitons in the different layers of the heterostructures is another pathway towards coherent control for quantum information applications.
For feasibility, identifying areas with robustness of these properties - charge transfer, ILE lifetimes, and coherent coupling strength - is thus a requirement.

However, the intricate spatial and temporal variations of exciton interactions with their environment and amongst themselves in TMD monolayers and heterostructures have remained mostly elusive due to limitations in the employed experimental techniques.
Here we use multi-dimensional coherent imaging spectroscopy (MDCIS) to map the distribution of dephasing times and inhomogeneity across an MoSe\textsubscript{2} monolayer, enabling the identification of promising and unfavorable areas for quantum information applications.
MDCIS also gives insight into strain across the monolayer. Correlating it with findings from Photoluminescence (PL)-spectroscopy and PL-detected MDCIS, we visualize the distribution of bright and dark exciton states across the monolayer sample. 
We further use MDCIS to spatially map strain and conduction band changes across a MoSe\textsubscript{2}/WSe\textsubscript{2} heterostructure. Despite the presence of notable, complex local strain and dielectric environment changes, we demonstrate surprising robustness of key sample properties: The rapid electron and hole transfer as well as coherent coupling between intralayer excitons in the MoSe\textsubscript{2} and WSe\textsubscript{2} monolayers and interlayer exciton lifetimes are robust across a majority of the heterostructure.
These results have larger implications for the commonly made device application claims, strengthening the case for TMDs as the materials of choice in various applications, including quantum information.

\section{Experimental Methods}

\subsection*{Multi-dimensional coherent spectroscopy (MDCS)}

The multi-dimensional coherent spectroscopy (MDCS) technique that is the basis for the majority of results in this paper is schematically shown in Fig.\,\ref{fig:Fig1}(a). MDCS uses a four-pulse sequence where the first two pulses (A,B) act as the pump pulse, while the third pulse (C) acts as a probe pulse and a fourth pulse (D) is used for heterodyne detection.
The first pulse (A) excites a coherence between the ground state and excited state, while the second pulse (B) converts this into an excited or ground state population. The second pulse can also convert the system into a Raman-like non-radiative coherence, between different excited states of the excitons, that oscillates during $T$. The third pulse (C) reverts the population or non-radiative coherence back into a radiative coherence between ground and excited state. Because MDCS measures the phase-resolved response, scanning the time delay $\tau$ between the first two pulses gives access to the absorption energy upon Fourier transform while scanning the time delay $t$ between C and D gives access to the emission energy axis. Additional dynamics, such as population decay, charge transfer, and coherent coupling are accessible via the $T$ time delay.

To be compatible with diffraction-limited imaging, all four pulses need to be collinear, that is, all excitations and the signal must be overlapped in the same beam. To extract the signal against a large background, each pulse is frequency tagged with an acousto-optic modulator \cite{Eric_TMD,CoherentCouplingPurz,Tekavec} and the interference between the third-order nonlinear response (induced by the first three pulses) and the fourth pulse can subsequently be phase-sensitively detected using a lock-in amplifier. The absorption of beam D can be taken into account by measuring the reflected spectrum using the interference between C and D.

\subsection*{Multi-dimensional coherent imaging spectroscopy (MDCIS)}

In MDCIS, instead of taking a single data point for fixed $\tau$, $T$, and $t$ positions, an entire image of the sample is taken. This yields an up to five-dimensional data set with full temporal information along $\tau$, $T$, and $t$ for every image pixel. Subsequent Fourier transforms along the $\tau$ and $t$ axes leads to an MDCS spectrum at every single image pixel. This process is illustrated for a hexagonal Boron Nitride (hBN) encapsulated MoSe\textsubscript{2} monolayer [Fig.\,\ref{fig:Fig1}(b)] in Fig.\,\ref{fig:Fig1}(c). MDCIS enables the study of linewidth evolutions across the sample as well as coupling dynamics by harnessing the rich spectroscopic and temporal information of MDCS and the rich spatial information of nonlinear imaging. 
The imaging is implemented with a laser-scanning imaging setup shown in Fig.\,\ref{fig:Fig1}(d), allowing for a diffraction-limited beam diameter (Abbe limit) of 940\,nm.
Because of the third-order nonlinearity of the four-wave mixing (FWM) based MDCS, the spatial resolution for the FWM measurements reaches 540\,nm. The four pulses impinge on an $x$- and $y$-Galvo mirror, with the $x$-Galvo mirror being relayed onto the $y$-Galvo mirror via a 4$f$-setup incorporating off-axis parabolic mirrors. The deflected beams are subsequently sent through another 4$f$-setup containing a scan and tube lens optimized for broadband, wide field-of-view laser scanning microscopy. This arrangement images the angle of the laser beam onto the objective without changing the position on the objective aperture, allowing for aberration-free laser scanning imaging. We employ a custom lock-in amplifier \cite{OPL} for rapid imaging.
A Ti:Sapph laser with nearly transform-limited 35\,fs pulses is used in this work. The pulses are pre-compensated with negative dispersion using an SLM-based pulse shaper to compensate for the dispersion acquired upon propagation through the setup. All experiments are performed at 6\,K.

\subsection*{Sample fabrication}

The two samples studied in this work are an hBN encapsulated MoSe\textsubscript{2} monolayer and an hBN encapsulated MoSe\textsubscript{2}/WSe\textsubscript{2} heterostructure. The samples were assembled using a dry-transfer technique with a stamp made of a polydimethylsiloxane cylinder with a thin film of poly (bisphenol A carbonate) on top.

\section{Results and Discussion}

\subsection{MoSe\textsubscript{2} monolayer}

The white light microscopy image of the MoSe\textsubscript{2} monolayer in Fig.\,\ref{fig:Fig1}(b) shows a region with cracks and bubbles towards the center but otherwise appears mostly pristine. The upper-left part of the monolayer region (shaded) is non-encapsulated, while the lower-right side of the monolayer (unshaded) is encapsulated. Given the reported instabilities and inconsistencies of non-encapsulated samples \cite{Kasprzak,Eric_TMD,Kasprzak_LW2}, we will focus our studies on the encapsulated area of the sample.
An integrated four-wave mixing (FWM) image of the MoSe\textsubscript{2} monolayer is shown in Fig.\,\ref{fig:Fig1}(e).
A large area on the top shows a very strong and approximately homogeneous FWM amplitude, while a dark area towards the center of the samples coincides with cracks and bubbles visible in the white light microscope image in Fig.\,\ref{fig:Fig1}(b) The bottom area of the sample again shows a comparatively homogeneous strength.
This observation is in contrast to photoluminescence (PL)-spectroscopy measurements plotted in Fig.\,\ref{fig:Fig1}(f). Here, the integrated PL strength for the center area (yellow square) is slightly higher than for the homogeneous area of the sample (green square). This observation is further supported by PL-detected FWM imaging (see Supplementary Information). As discussed by Smallwood \textit{et al.} \cite{Chris_PRL}, this suggests a highly spatially dependent distribution of bright and dark states. 
Furthermore, a noticeable linewidth change between the center and edge area of the monolayer can be observed in the PL spectra.
This linewidth observation is corroborated by the exemplary MDCS spectra for the homogeneous bottom area of the sample [Fig.\,\ref{fig:Fig1}(g)] and the center, dark area [Fig.\,\ref{fig:Fig1}(h)]. The center area shows a more dominant inhomogeneous broadening, which manifests itself as an elongation along the diagonal (dashed line), compared to the bottom sample area. The broader linewidth in the PL spectrum in Fig.\,\ref{fig:Fig1}(f) is thus caused by an increase in inhomogeneity, and not related to homogeneous linewidth changes.
Moreover, the inhomogeneously broadened spectrum on the center area shows more weight towards lower energies, suggesting the existence of notable resonance shifts on the order of several meV. 

To gain a better understanding of the underlying physics and identify promising areas for quantum information applications, a more systematic study of resonance shifts and linewidths is required.

As pointed out by Jakubczyk\cite{Kasprzak_LW2,Kasprzak_LW1} and Boule \textit{et al.} \cite{Kasprzak_Coherent}, there is an expected anti-correlation between the homogeneous linewidth $\gamma$ and inhomogeneous linewidth $\sigma$. There is also an expected anti-correlation between $\sigma$ and the dipole moment $\mu$, and hence the FWM strength ($\propto \mu^4$). 
In the presence of smaller $\sigma$, the exciton coherence volume in real space is larger. This increases the light matter interaction because of the smaller $k$-space volume and larger overlap with the light-cone, increasing the dipole moment $\mu$ and decreasing the radiative lifetime $T_{\mathrm{rad}}$. Thus, a small inhomogeneity $\sigma$ generally favors higher $\gamma$ and, via $T_{2}=\hbar/\gamma$, smaller dephasing times. However, in the presence of non-radiative homogeneous broadening mechanisms such as non-radiative electron scattering, $\gamma$ can be further increased \cite{Kasprzak_LW1,Kasprzak_LW2}. For quantum information applications, a small $\gamma$ and $\sigma$ is desired.

\begin{figure}[t]
\centering
\includegraphics[width=0.5\textwidth]{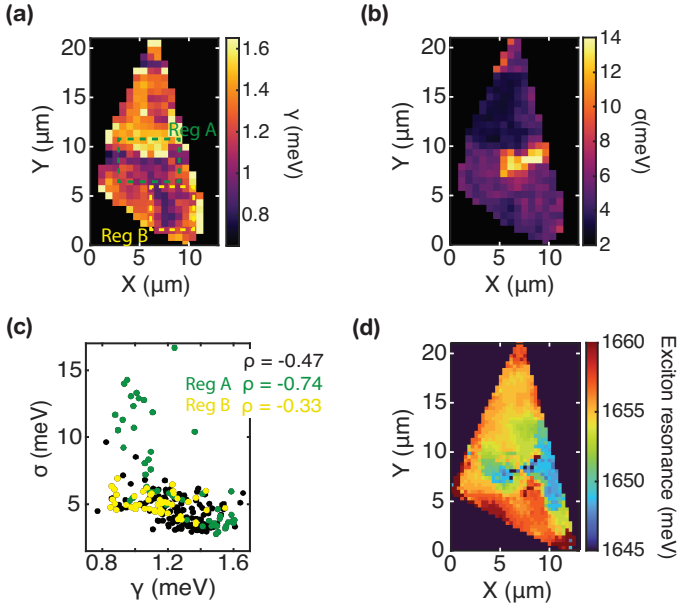}
\caption{\label{fig:Fig2} \textbf{(a,b)} Homogeneous ($\gamma$) and inhomogeneous linewidth ($\sigma$) map of the MoSe\textsubscript{2} monolayer. We average 2x2 pixels together for these maps, obtaining a pixel size close to the spatial resolution. For legibility, the color bar in (b) is capped at 14\,meV. \textbf{(c)}  $\sigma$ vs. $\gamma$ for all sample points. Sample points falling within the area of the green (yellow) rectangle in (a) are plotted in green (yellow), all other sample points are plotted in black. Correlation for the sample points within the colored rectangles and all data points together is calculated using the Pearson correlation coefficient. \textbf{(d)} Resonance energy map of the MoSe\textsubscript{2} monolayer.}
\end{figure}

Maps of $\gamma$ and $\sigma$ across the MoSe\textsubscript{2} monolayer are shown in Fig.\,\ref{fig:Fig2}(a) and (b) respectively. The linewidths are extracted by simultaneously fitting diagonal and cross-diagonal slices in the 2D frequency data with the analytical solutions for abitrary amounts of $\gamma$ and $\sigma$ provided in \cite{Fits}. This is notably different from the procedure employed in \cite{Kasprzak_LW2,Kasprzak_LW1,Kasprzak_Coherent}, where linewidths are fitted in the temporal domain. As discussed in the Supplementary Information, the frequency domain is less susceptible to measurement noise because of the Fourier transform filtering out a majority of the high frequency noise. Furthermore, the frequency domain fitting procedure emphasizes the high signal areas, as elucidated upon in the Supplementary Information. After fitting, we further average $2 \times 2$ pixels together to reduce pixel-to-pixel noise. The averaging leaves us with a pixel size of 700\,nm, close to the spatial resolution.

The areas of lower FWM show a significantly increased $\sigma$ and a smaller $\gamma$. Overall, $\gamma$ shows drastic changes, ranging between 0.75\,meV and 1.7\,meV, more than a factor of two difference. Depending on the sample spot $\sigma$ increases by up to a factor of 10, from below 3\,meV to values up to 35\,meV. Linewidths changes in conjunction with dipole changes are usually assigned to strain, with high strain areas showing large $\sigma$, smaller $\gamma$, and smaller $\mu$ \cite{Kasprzak_LW1,Kasprzak_LW2}.

The anti-correlation between $\gamma$ and $\sigma$ across large areas of the sample can best be visualized by plotting $\sigma$ against $\gamma$, as done in Fig.\,\ref{fig:Fig2}(c). Here, sample points falling within the green and yellow rectangle [Fig.\,\ref{fig:Fig2}(a)] are plotted in green and yellow respectively, while all other sample points are plotted in black.
A moderately strong anti-correlation of $\rho=-0.47\pm0.1$ (Pearson correlation coefficient) can be observed for the two linewidths across the entirety of the sample.
A closer inspection of Fig.\,\ref{fig:Fig2}(c) together with Fig.\,\ref{fig:Fig2}(a,b) suggests that there are areas of stronger anti-correlation, while other areas show a weaker correlation between the linewidths. This is further supported by plotting the two areas marked with a green and yellow rectangle in Fig.\,\ref{fig:Fig2}(c). Indeed, the green sample points show a strong anti-correlation with a correlation coefficient $\rho=-0.74\pm0.13$, while the yellow points shows a much weaker anti-correlation with a correlation coefficient $\rho=-0.33\pm0.26$. Most importantly with regards to quantum information applications,  the yellow points show values of $\gamma$ around 0.9\,meV  with comparatively low $\sigma$ between 4.5-7\,meV, while for the green rectangle similar values of $\gamma$ can only be found with $\sigma$ above 10\,meV. Hence, the yellow rectangle area towards the bottom of the sample is a more favorable area than the green rectangle for quantum information applications that rely on low inhomogeneity and slow dephasing. However, another low strain area towards the top of the sample, while showing overall lower $\sigma$ also shows higher $\gamma$, illustrating that a low strain area is not inherently a "good" area of the sample with regards to the application potential. With additional non-radiative contributions to $\gamma$ such as non-radiative electron scattering \cite{Kasprzak_LW1,Kasprzak_LW2}, affected for example by doping, only a technique like MDCIS allows to unambiguously identify the relevant sample areas.

Assigning the linewidth and dipole changes to strain, we can further distinguish between tensile and compressive strain via the sign of resonance energy shifts across the sample \cite{ChernikovDielectric}.
We plot the exciton resonance energy across the sample in Fig.\,\ref{fig:Fig2}(d).
This map shows significant shifts of the resonance energy between 1645\,meV and up to 1660\,meV across the 10$\times$15\,${\mu \mathrm{m}}^2$ area of the sample. The areas of strongest resonance shifts also coincide with the areas of lower integrated FWM.  The observation of both red and blue shifts across the sample points towards complex local strain dynamics involving both tensile (lower energies) and compressive (higher energies) strain, caused by the cracks and bubbles in the central area of the sample. 

Having established overall strain dynamics and how intralayer excitons are affected in MDCIS across an encapsulated MoSe\textsubscript{2} monolayer, we can take these findings to a more complicated material system, namely a MoSe\textsubscript{2}/WSe\textsubscript{2} heterostructure.

\begin{figure*}
\centering
\includegraphics[width=1.0\textwidth]{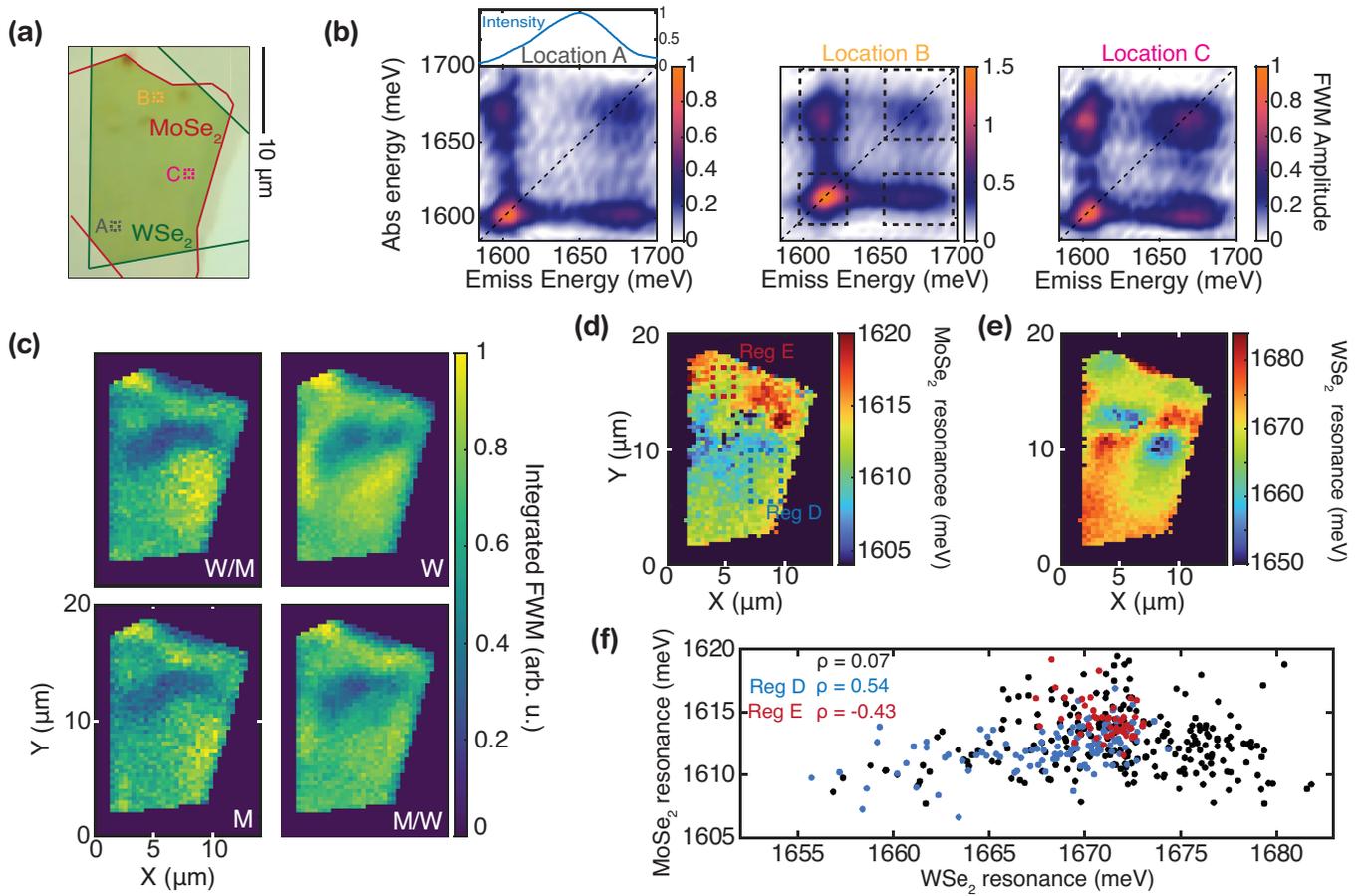}
\caption{\label{fig:Fig3} \textbf{(a)} White light microscopy image of the hBN-encapsulated MoSe\textsubscript{2}/WSe\textsubscript{2} heterostructure. \textbf{(b)} Low-power, low-temperature MDCS spectra of MoSe\textsubscript{2}/WSe\textsubscript{2} heterostructure at three different sample points marked by the three squares in (a). The laser spectrum used for all MDCIS experiments on the heterostructure is plotted atop. \textbf{(c)} Integrated FWM of the four peaks shown in (b). The integration area for the four peaks is shown for Location B in (b) and is kept fixed across the sample. The area of low intensity is a high strain area with cracks and wrinkles. \textbf{(d)} MoSe\textsubscript{2} resonance energy map across the MoSe\textsubscript{2}/WSe\textsubscript{2} heterostructure. \textbf{(e)} WSe\textsubscript{2} resonance energy map across the MoSe\textsubscript{2}/WSe\textsubscript{2} heterostructure. \textbf{(f)} MoSe\textsubscript{2} resonance energy vs. WSe\textsubscript{2} resonance energy for all sample points. Sample points falling within the area of the red (blue) rectangle in (d) are plotted in red (blue), all other sample points are plotted in black. Correlation for the sample points within the colored rectangles and all data points together is calculated using the Pearson correlation coefficient.}
\end{figure*}

\subsection{MoSe\textsubscript{2}/WSe\textsubscript{2} heterostructure}

We extend our study to a MoSe\textsubscript{2}/WSe\textsubscript{2} heterostructure encapsulated in hBN.
Fig.\,\ref{fig:Fig3}(a) shows a white light microscopy image of the heterostructure. Three exemplary MDCS spectra taken at the three points marked in Fig.\,\ref{fig:Fig3}(a) are plotted in Fig.\,\ref{fig:Fig3}(b). The two on-diagonal (dashed line) peaks are associated with the MoSe\textsubscript{2} and WSe\textsubscript{2} intralayer A-exciton. The two off-diagonal peaks are indicative of both coherent coupling and incoherent electron and hole (charge) transfer, as discussed in previous work \cite{CoherentCouplingPurz}. The spectra show significant energy shifts and varying peak strengths for both the MoSe\textsubscript{2} and WSe\textsubscript{2} resonances across the sample.
We spectrally integrate over the four peaks to better visualize their strength variations across the sample. Maps of the integrated FWM for the four peaks are shown in Fig.\,\ref{fig:Fig3}(c). The figures have the same order as the peaks - with the MoSe\textsubscript{2} (M) peak in the lower left, the WSe\textsubscript{2} (W) peak in the upper right, and the MoSe\textsubscript{2}/WSe\textsubscript{2} (M/W) and WSe\textsubscript{2}/MoSe\textsubscript{2} (W/M) peak in the lower right and upper left respectively.
Similar to the MoSe\textsubscript{2} monolayer, there is a region of decreased FWM towards the upper center of the sample, which is associated with a high strain area due to wrinkles and bubbles that are caused by fabrication. Apart from this area, the peak strength is mostly homogeneous across the sample for all four peaks, except for the MoSe\textsubscript{2} and MoSe\textsubscript{2}/WSe\textsubscript{2} peaks having a lower strength towards the bottom left of the sample. The differences in peak strengths can partially be attributed to spatial variations of the dipole moment $\mu$ that differ for the two materials, based on the local strain profile. Furthermore, the finite bandwidth of the employed laser and the reduced excitation density at sample points where the WSe\textsubscript{2} (MoSe\textsubscript{2}) resonance is shifted to higher (lower) energies can contribute to the spatial peak strength variations. 

The resonance energy shifts for the two resonances are plotted in Fig.\,\ref{fig:Fig3}(d,e). A large shift towards lower energies down to 1605\,meV for the MoSe\textsubscript{2} resonance can be observed in the center-left of the sample, while the WSe\textsubscript{2}  resonance shifts towards higher energies, up to 1680\,meV in this area. Towards the bottom area of the sample, both resonances shift towards higher energies. Moreover, the MoSe\textsubscript{2} resonance shifts towards significantly higher energies up to 1620\,meV in the upper part of the sample. To correlate the resonance shifts, we plot the MoSe\textsubscript{2} resonance vs. the WSe\textsubscript{2} resonance energy in Fig.\,\ref{fig:Fig3}(f). Sample points falling within the red (blue) rectangle are plotted in red (blue), all other data points are plotted in black. While changes in the resonance energies seem to be correlated, a Pearson coefficient of $\rho=0.07\pm0.13$ suggests otherwise. A closer examination of Fig.\,\ref{fig:Fig3}(f) together with Fig.\,\ref{fig:Fig3}(d) and (e) shows both strong correlation and anti-correlation, which cannot be captured by the Pearson correlation coefficient. 
Instead, to visualize the strong variation of correlation across the sample, we examine the select areas marked by a blue and red dashed rectangle in Fig.\,\ref{fig:Fig3}(d).
For the area marked by the blue rectangle (Reg D), a strong correlation with $\rho=0.54\pm0.12$ can be observed. A correlation between the two resonances is what would be expected: It is well known that both encapsulation and heterostructure formation commonly redshift the excitons \cite{PasqualILE,Eric_TMD}, while compressive (tensile) strain leads to red (blue) shifts \cite{Khatibi_Strain}. Both monolayers should be affected in the same way by these factors. Contrary to this expectation, the area marked by the red rectangle (Reg E) shows a moderately strong anti-correlation with $\rho=-0.43\pm0.22$. This points towards complex local strain dynamics, where the two monolayers experience different strain that leads to opposite resonance shifts. One possible scenario is the compression of one monolayer which leads to a bubble in the other monolayer, inducing tensile strain and thus the opposite sign resonance shift.
The attribution to strain is further supported by the fact that the largest resonance shifts can again be observed around an area of low FWM with physical imperfections (cracks, bubbles, wrinkles) visible in white light microscopy [see Fig.\,\ref{fig:Fig3}(a)].

\begin{figure}
\centering
\includegraphics[width=0.5\textwidth]{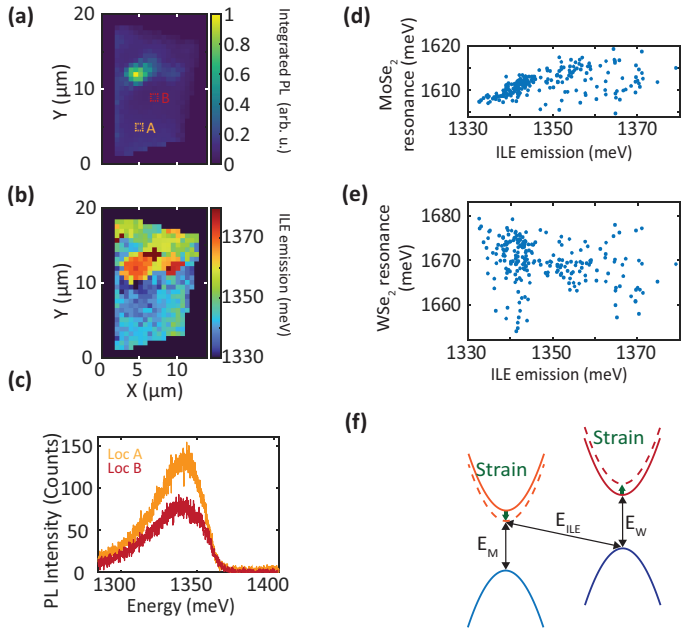}
\caption{\label{fig:Fig4} \textbf{(a)} Integrated interlayer exciton PL across the sample. \textbf{(b)} Map of interlayer exciton PL-emission energy \textbf{(c)} PL-spectrum for two points on center and bottom area of the heterostructure marked in (a). \textbf{(d)} Correlation between MoSe\textsubscript{2} resonance energy and interlayer exciton emission energy. \textbf{(e)} Correlation between WSe\textsubscript{2} resonance energy and interlayer exciton emission energy. \textbf{(f)} Schematic of conduction and valence band for MoSe\textsubscript{2} and WSe\textsubscript{2} under the influence of strain.}
\end{figure}

The local strain profile also leads to large variations in the interlayer exciton (ILE) PL. The integrated ILE PL plotted in Fig.\,\ref{fig:Fig4}(a) shows a strong maximum at the high-strain area of minimum FWM, but appears otherwise relatively homogeneous in strength across the sample. These observations align with the observations on the MoSe\textsubscript{2} monolayer.
The emission energy of the ILE is plotted in Fig.\,\ref{fig:Fig4}(b). The area of strong PL shows above average energy of ILE emission around 1370\,meV. The upper area of the sample also shows a higher ILE emission energy around 1355\,meV, while the lower part of the sample shows lower emission energies down to 1330\,meV. 
The strong changes of the ILE PL, including emission strength and energy, even outside the high strain area, can be further emphasized when plotting two select spectra in what appears to be a relatively homogeneous region in the bottom half of the sample. These spectra plotted in Fig.\,\ref{fig:Fig4}(c) demonstrate that even in the bottom part of the sample, ILE emission is highly spatially heterogeneous.

We further observe a strong correlation between the ILE PL-emission energy and the MoSe\textsubscript{2} resonance energy (measured with MDCS), as evident from Fig.\,\ref{fig:Fig4}(d) and a correlation coefficient $\rho=0.4705$. The correlation with the WSe\textsubscript{2} resonance [Fig.\,\ref{fig:Fig4}(e)] is much weaker ($\rho=-0.2723$). This difference can be explained as follows: As Khatibi \textit{et al.}\cite{Khatibi_Strain} show for TMD monolayers, along the K-point, mainly the conduction band is affected by strain, while effects on the valence band are negligible. Thus, as illustrated in Fig.\,\ref{fig:Fig4}(f), the strain-induced conduction band shift in the MoSe\textsubscript{2} immediately affects the ILE emission energy, while the strain-induced shift in the WSe\textsubscript{2} should have no effect. The residual anti-correlation between ILE emission energy and WSe\textsubscript{2} resonance energy stems from the local (anti-)correlation between MoSe\textsubscript{2} and WSe\textsubscript{2} established in the discussion above.
Given these significant strain-induced changes to the heterostructure, if, and how much these changes affect the coupling between excitons in the two layers is a topic of interest.

\begin{figure*}[t]
\centering
\includegraphics[width=1.0\textwidth]{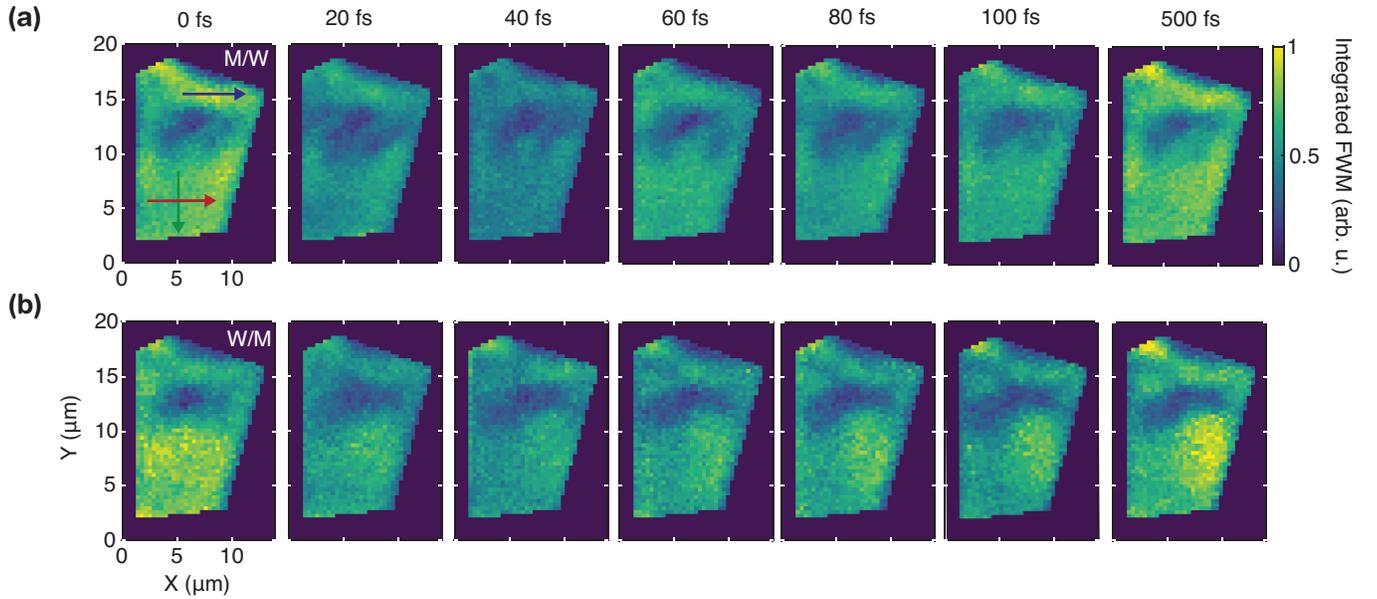}
\caption{\label{fig:Fig5} \textbf{(a)} Integrated FWM of the MoSe\textsubscript{2}/WSe\textsubscript{2} coupling peak for varying $T$ delays. \textbf{(b)} Integrated FWM of the WSe\textsubscript{2}/MoSe\textsubscript{2} coupling peak for varying $T$ delays. Both peaks are normalized to their respective maximum at T=0\,fs. The strong amplitude signatures of the coherent coupling oscillations and charge transfer are evident.}
\end{figure*}

\begin{figure*}[ht]
\centering
\includegraphics[width=1.0\textwidth]{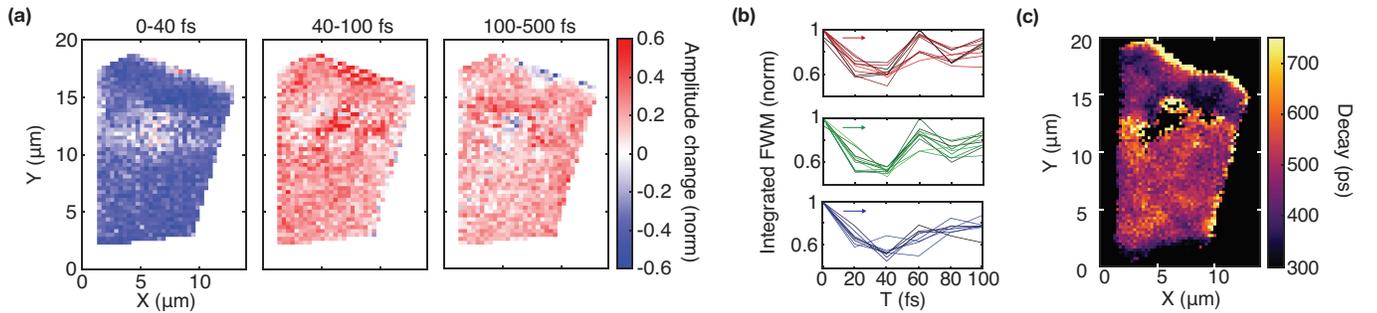}
\caption{\label{fig:Fig6} \textbf{(a)} Difference in integrated FWM amplitude for the MoSe\textsubscript{2}/WSe\textsubscript{2} peak between various $T$ delays, normalized by the MoSe\textsubscript{2}/WSe\textsubscript{2} peak amplitude at $T$=0\,fs. This visualization emphasizes the homogeneity of the coherent coupling and charge transfer. \textbf{(b)} Integrated FWM of the MoSe\textsubscript{2}/WSe\textsubscript{2} coupling peak for varying $T$ delays along the arrows indicated in Fig.\,\ref{fig:Fig5}(c). \textbf{(c)} Decay time ($T$) map of the FWM signal taken at $t=\tau=0$.}
\end{figure*}

To shed some light on the coupling dynamics, we employ dynamic MDCIS, where the pump-probe delay $T$ [see Fig.\,\ref{fig:Fig1}(a)] is varied to access coherent and incoherent coupling dynamics in the heterostructure. We plot the integrated FWM of the MoSe\textsubscript{2}/WSe\textsubscript{2} (M/W) and WSe\textsubscript{2}/MoSe\textsubscript{2} (W/M) coupling peaks for varying $T$ delay in Fig.\,\ref{fig:Fig5}(a) and (b) respectively. 
Initially, the signal decays for both coupling peaks from 0 to 40\,fs but recovers again for 60\,fs. A smaller variation of the integrated amplitudes can be observed between 60-100\,fs. This behavior can be assigned to coherent coupling oscillations, that have been shown to occur in MoSe\textsubscript{2}/WSe\textsubscript{2} heterostructures \cite{CoherentCouplingPurz}. The coherent oscillation has an amplitude of approximately 50\% (with respect to the peak amplitude) across the sample, the signature of strong coupling between excitons in the two layers. As discussed by Hao \textit{et al.} \cite{CoherentGalan}, the amplitude is below unity due to interference between different coherent coupling contributions and exponentially decaying phase-space filling nonlinearities.
In between $T$=100\,fs and $T$=500\,fs both coupling peaks show a clear rise that we assign to the charge transfer that has been observed using MDCS in this specific heterostructure \cite{CoherentCouplingPurz}, as well as other TMD heterostructures \cite{MDCS_CT}. While not resolved here, the charge transfer for the WSe\textsubscript{2}/MoSe\textsubscript{2} peak is faster in this sample, as indicated by both previous measurements \cite{CoherentCouplingPurz}, as well as the broader linewidths of the WSe\textsubscript{2} peak in Fig.\,\ref{fig:Fig3}(b).
The strong amplitude change between 100\,fs and 500\,fs suggests highly efficient electron and hole transfer in the heterostructure. 
Most strikingly though, both coherent coupling and charge transfer appear mostly homogeneous across the sample. This is evident from the fact that the relative peak strength across the sample remains mostly unchanged along $T$, while the absolute peak strength changes. The homogeneity can be visualized by comparing the relative strength profiles at $T$=0\,fs, $T$=40\,fs, $T$=100\,fs, and $T$=500\,fs, as done in Fig.\,\ref{fig:Fig6}(a). Here, we plot the difference between the integrated FWM maps for the M/W peak, normalized by the M/W map at $T$=0\,fs. We plot the difference between $T$=0\,fs and $T$=40\,fs (leftmost map), $T$=40\,fs and $T$=100\,fs (center map), and $T$=100\,fs and $T$=500\,fs (right map). Both the initial decay and recovery, part of the coherent coupling oscillations, show a homogeneous strength profile across the entire sample. Moreover, the charge transfer rise between 100\,fs and 500\,fs shows the same homogeneous strength profile.

Moderate resonance shifts due to strain on the order of 10-20\,meV are not inherently expected to change the charge transfer dynamics significantly because the shift is comparatively small to the band offsets of hundreds of meVs \cite{Bandoffset1,Bandoffset2}. However, the sensitivity of charge transfer to sample parameters such as the twist angle \cite{CT_twistangle,CoherentCouplingPurz} and lattice separation \cite{CT_distance} is well documented. With the resonance shifts being an indicator of complex local strain dynamics, this strain is expected to change the interlayer spacing, among other things. These previous findings render our observation of spatially homogeneous charge transfer rather surprising.
Moreover, the coherent coupling between the resonances is expected to be equally dependent on the intralayer separation.
Indeed, a limited decrease of the coherent coupling strength is observable for the top part of the sample, as illustrated in Fig.\,\ref{fig:Fig6}(b). Here, we plot the integrated FWM of the MoSe\textsubscript{2}/WSe\textsubscript{2} coupling peak along the three arrows drawn in Fig.\,\ref{fig:Fig5}(a). Both the green and red arrows, on the lower part of the sample, show a strong coherent coupling oscillation with only small, random changes when moving across the sample, consistent with the observations in Fig.\,\ref{fig:Fig6}(a). However, the data for the blue arrow shows a weaker oscillation, which might be either caused by a reduced amplitude or a more rapid dephasing of the coherent coupling in the upper area of the sample. 
Given the strong resonance shifts for the MoSe\textsubscript{2}resonance in this area as well as the reflectance map (see supplemental information) we assume this to be an area of increased interlayer spacing. The increased interlayer spacing can explain the reduced coherent coupling strength/more rapid dephasing.

This specific area also shows up distinctly in the FWM decay time map plotted in Fig.\,\ref{fig:Fig6}(c). The decay time map is acquired at $t=\tau=0$ while moving the $T$ delay [see Fig.\,\ref{fig:Fig1}(a)].
The FWM decay in the heterostructure for $T$>50\,ps is dominated by interlayer exciton decay through ground-state bleaching contributions to the signal \cite{MDCS_review}. The decay time can thus be taken as an indirect probe of the interlayer exciton lifetime. The bottom area of the sample again shows no notable spatial inhomogeneity in the lifetime, with values in the range between 500-550\,ps. Given the low twist angle of this sample, these values are in good agreement with the literature \cite{ILE_twistAngle_lifetime}. 
The upper area of the sample has on average a 30-40\% lower lifetime and larger inhomogeneity with values ranging from 300-550\,ps.
The combination of reduced interlayer exciton lifetime as well as reduced coherent coupling, together with the resonance shifts observed in this area, show that the sample properties are not entirely immune to strain and defect induced changes. Nonetheless, the relatively low sensitivity towards these changes remains surprising.

\section{Conclusions and Outlook}

We have examined an encapsulated MoSe\textsubscript{2} monolayer and an encapsulated MoSe\textsubscript{2}/WSe\textsubscript{2} heterostructure using multi-dimensional coherent imaging spectroscopy. We image strain, dephasing, and inhomogeneity across the monolayer, showing a moderately strong anti-correlation between homogeneous and inhomogeneous linewidths that increases and decreases based on the area on the sample. Using MDCIS allows us to unambiguously identify promising and unfavorable areas for quantum information applications.
We also visualize a spatially dependent dark vs. bright state distribution using the difference between heterodyne detected FWM images, PL, and PL-detected four-wave mixing.
Carrying this technique over to the heterostructure, we map the complex strain dynamics by correlating resonance energies. Our observations lead us to propose a phenomenological model for the band structure changes due to strain variations, using additional interlayer exciton photoluminescence measurements.
Employing dynamic multi-dimensional coherent imaging spectroscopy, we visualize coherent coupling and electron and hole transfer across the sample.
While certain areas of the sample show a reduced coherent coupling and decreased interlayer exciton lifetime, we demonstrate an overall robustness of the coupling dynamics to strain and defects across the sample.

The future of these materials, especially in the realm of quantum information, is inherently coupled to the scalability and quality of fabricated devices. 
This work shows a reproducibility of crucial physical properties - dephasing time, coupling strength, and interlayer exciton lifetime - across large areas of the sample. The reproducibility lays the groundwork and strengthens the case for transition metal dichalcogenides as a next generation material.
We further demonstrate the usefulness of the multi-dimensional coherent imaging spectroscopy technique which, on a smaller scale, has been realized by spatially-addressed multi-dimensional coherent spectroscopy \cite{Kasprzak_Coherent,Kasprzak_LW1,Kasprzak_LW2,Ogilvie}. However, recent technological advances in lock-in detection \cite{OPL} allow for a larger scale realization of multi-dimensional coherent imaging spectroscopy and related techniques, advancing the forefront of materials and device characterization.

\section*{Supplementary Material}

See supplementary material for additional information on the experimental setup, the linewidth fitting procedure, reflection maps of the samples, and PL-detected FWM imaging data on the MoSe\textsubscript{2} monolayer.

\begin{acknowledgments}
We thank Blake Hipsley for the construction of the pulse shaping setup.
The research at University of Michigan was supported by NSF Grant No. 2016356. The work at University of Washington is supported by the Department of Energy, Basic Energy Sciences, Materials Sciences and Engineering Division (DE-SC0012509). W.G.H. was supported by the NSF Graduate Research Fellowship Program under Grant No. DGE-1762114.
\end{acknowledgments}

\section*{Author Declarations}
\subsection*{Conflict of interest}

EWM and STC are co-founders of MONSTR Sense Technologies, LLC, which sells ultrafast spectrometers and microscopes.

\section*{Data Availability Statement}

The data that support the findings of this study are available from the corresponding author upon reasonable request.

\bibliography{aipsamp}

\end{document}